\documentclass[iop]{emulateapj}

\begin{document}

\title{Oblique Shocks As The Origin Of Radio To Gamma-ray Variability In AGN}
\author{Philip A. Hughes, Margo F. Aller, and Hugh D. Aller}
\affil{Astronomy Department, University of Michigan, Ann Arbor, MI 48109-1042}
\email{phughes@umich.edu, mfa@umich.edu, haller@umich.edu}

\begin{abstract}
The `shock in jet' model for cm-waveband blazar variability is revisited,
allowing for arbitrary shock orientation with respect to the jet flow
direction, and both random and ordered magnetic field. It is shown that
oblique shocks can explain events with swings in polarization position
angle much less than the $90\arcdeg$ associated with transverse structures,
while retaining the general characteristics of outbursts, including spectral
behavior and level of peak percentage polarization. Models dominated by
a force-free, minimum energy magnetic field configuration (essentially
helical) display a shallow rise in percentage polarization and frequency
dependent swing in polarization position angle not in agreement with the
results of single-dish monitoring observations, implying that the field
is predominantly random in the quiescent state. Outbursts well-explained
by the `shock in jet' model are present during $\gamma$-ray flaring in
several sources, supporting the idea that shock events are responsible
for activity from the radio to $\gamma$-ray bands.
\end{abstract}

\section{Introduction}\label{intro}
One of the early results from EGRET was the discovery that a subset of very
active extragalactic objects, blazars, emit at $\gamma$-ray energies. A close
association between activity in the radio band identified using total flux
density monitoring observations and detections by EGRET in the GeV band was
established in the mid-1990s, e.g.,  \citet{val95}, an association which
has been confirmed using the large data base of measurements provided by
{\it Fermi} and monitoring measurements from both single dish and VLBA
imaging measurements, e.g., \citet{kov09,ric10}.  For the past 25 years,
the accepted explanation for blazar variability in the optical-to-radio
bands has been the `shock in jet' model; hence the temporal associations
between the activity in the radio and $\gamma$-ray bands and the fact
that the EGRET detections occurred during the rise portions of the radio
flares suggested to investigators that the same shocks producing the radio
flares were responsible for the emission in the $\gamma$-ray band, e.g.,
\citet{val96}.  This shock scenario has been echoed in recent studies of
{\it Fermi} observations of AGN, e.g., \citet{abd10b}, but rigorous tests
have not been extensively carried out.

Evidence in support of the `shock in jet' scenario originally came from
model fits to the broadband spectral evolution in the optical-to-radio
band in 3C 273 \citep{mar85} and independently from radiative transfer
model fits to multifrequency total flux density and linear polarization
monitoring measurements in the centimeter band \citep{hug89a,hug89b,hug91}.
The generally accepted scenario is that outbursts result from instabilities
which develop naturally within jet flows, producing shocks.  The magnetic
field is initially random within this emitting region \citep{jon85},
but the shocks produce a compression and an increased ordered component;
the expected signature in the radio band linear polarization light curves
for such a  shock event is a swing in the electric vector position angle
(EVPA; an orientation orthogonal to the magnetic field direction in a
transparent source) and an increase in the fractional linear polarization.
This early modeling assumed that the shocks had a specific orientation --
transverse to the flow direction. Attempts to fit later radioband events
in these same sources with the same model parameters used successfully
in the original fits, however, failed, and the characteristic behavior of
the variations suggested that shocks are more generally oriented obliquely
to the jet flow direction; thus, the initial transverse shocks identified
from their well-resolved, distinct appearance represented a special case
of a more general phenomenon.  Conical shocks have been discussed by
\citet{cc90}, and the formulation presented therein has been applied to
the data for some sources, e.g., \citet{mj10}.

Support for the shock model of major outbursts seen in single-dish data,
and the propagating components seen in maps of parsec-scale flows, plus
support for a shock explanation for at least some {\it Fermi} events,
would come from a) `revalidating' the `shock in jet' model, by showing that
oblique shocks can indeed explain the commonly observed reduced swing in EVPA
through only tens of degrees, and associated increases in both fractional
polarization and total flux density (flares) with the spectral behavior
exhibited in the data; b) showing that $\gamma$-ray flares occur at the
same time as radioband events plausibly explained by an oblique shock model.
It is also desirable to test whether the single-dish data can discriminate
between models involving purely random magnetic field, purely ordered
magnetic field, or a mixture of the two.  To those ends it is illustrated
here that the salient features identified in radio band linear polarization
and total flux density observations obtained by the University of Michigan
monitoring program (hereafter UMRAO) are reproduced by simulations using
plausible input parameters in modeling involving radiative transfer through
flows with propagating, oblique shocks; and it is demonstrated that this
shock signature is present during $\gamma$-ray flaring in several sources.

In \S\ref{umrao} the results from the Michigan variability program are
summarized, the general characteristics of the cm-waveband variability
are defined, and their relation to $\gamma$-ray flaring is discussed, with
examples. In \S\ref{flow} the details of an oblique shock model for this
cm-waveband activity are set out: the quiescent flow, models for both random
and ordered magnetic field components, the structure of an oblique shock,
and its propagation. The essential elements of an existing code for the
transfer of polarized radiation are recapped in \S\ref{transfer}, and the
results of applying this to a flow with representative flow parameters are
presented in \S\ref{results}: results for a range of observer orientations
and magnetic field topologies are contrasted. Section \ref{conclusions}
summarizes and discusses the findings that oblique shocks can explain the
observed cm-waveband events, the behavior of which support a model with
at most a modest contribution from an ordered magnetic field component.

\section{Observations}\label{umrao}
\subsection{The UMRAO Database}\label{umraodatabase}
As part of the Michigan variability program carried out with the 26-meter
radio telescope, source-integrated, single-dish total flux density
and linear polarization measurements have been obtained for hundreds of
sources, extending over time windows of up to 4 decades. The data acquired
provide a comprehensive view of the properties of blazar variability in
the centimeter band based on long term data acquired and reduced in a
consistent and uniform manner. These measurements commenced at 8~GHz in
the mid-1960s, at 14.5~GHz in 1974, and at 4.8~GHz in 1978. The number of
sources included in the UMRAO monitoring program increased dramatically
in the late 1970s when an automated observing program was implemented;
subsequently the fraction of time utilized per year for AGN observations
increased to over 90\%.  Typically 20-25 sources are observed in a 24 hour
period; each observation consists of a series of on-off measurements over
approximately 30-45 minutes. Measurements of calibrators are interspersed
with observations of program sources every 1 1/2 to 2 hours.  In general,
the observing cadence adopted for each program source is adjusted to
follow the variability in that particular AGN. Very active sources such as
those included in the shock program described here are typically observed
1-2 times per week at 14.5~GHz and once per week at 8 and 4.8~GHz during
flaring periods. This cadence is required to follow the rapid variations
in linear polarization expected from past observations.

\subsection{Characteristics Of The Variability}\label{variability}
Studies of the data for both individual sources and for statistical samples
have identified a number of general properties of the variability: the
characteristic variability time-scales of the total flux density, identified
from first-order structure analyses, are of the order of two years in the
observer's frame with some spread, e.g.,  \citet{hug92}. The ratios of
peak to quiescent amplitude of these variations are characteristically
only a factor of a few, reaching at most a factor of 8. While numerous
attempts have been made to identify periodic behavior in the total
flux density light curves, in general consecutive events do not repeat.
The outburst shapes are characterized by a rapid rise followed by a  more
gradual decline. In the millimeter band, the Mets\"ahovi monitoring data
have been successfully fit assuming a function with an exponential rise,
a sharp turnover, and an exponential decay with a time-scale approximately
1.3 times longer than the rise time-scale \citep{val99}, but a generic
shape is not apparent in the events observed in the lower frequency UMRAO
data, possibly due to the  blending of individual events.  The total
flux density spectra in the centimeter band are typically inverted and
relatively flat ($|\alpha|\leq0.5$); however, these differ from source to
source and can differ from event to event in the same source. The spectrum
itself in many cases evolves during an outburst; in many large events the
spectrum is inverted until burst maximum, and flattens during outburst
decline. Associated flares are often apparent in polarized flux density;
and in some cases individual events can be seen in linear polarization
which appear as blended events in total flux density.

\begin{figure}
\plotone{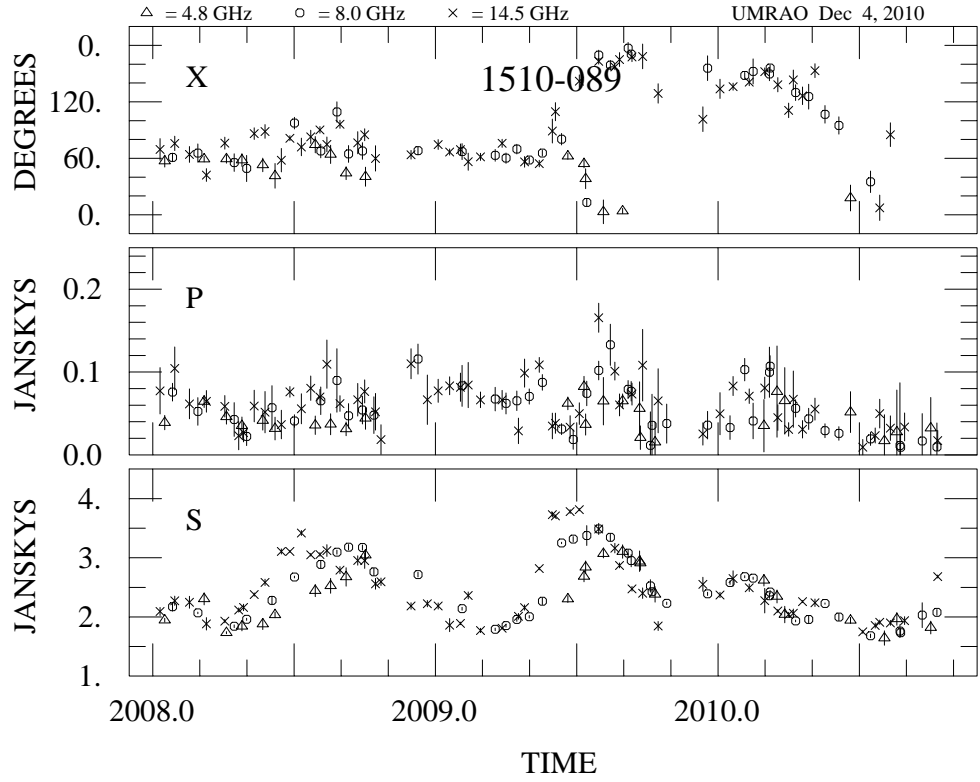}
\caption{From bottom to top: daily averages of the total flux density,
the linearly polarized flux density, and the electric vector position angle in
PKS~1510-089. The data at 14.5, 8.0 and 4.8~GHz are denoted by crosses,
circles and triangles respectively. The radio band flaring shown is
temporally associated with $\gamma$-ray activity. Note the differences in
spectral behavior during the flaring periods shown.}
\label{fig1}
\end{figure}

\begin{figure}
\plotone{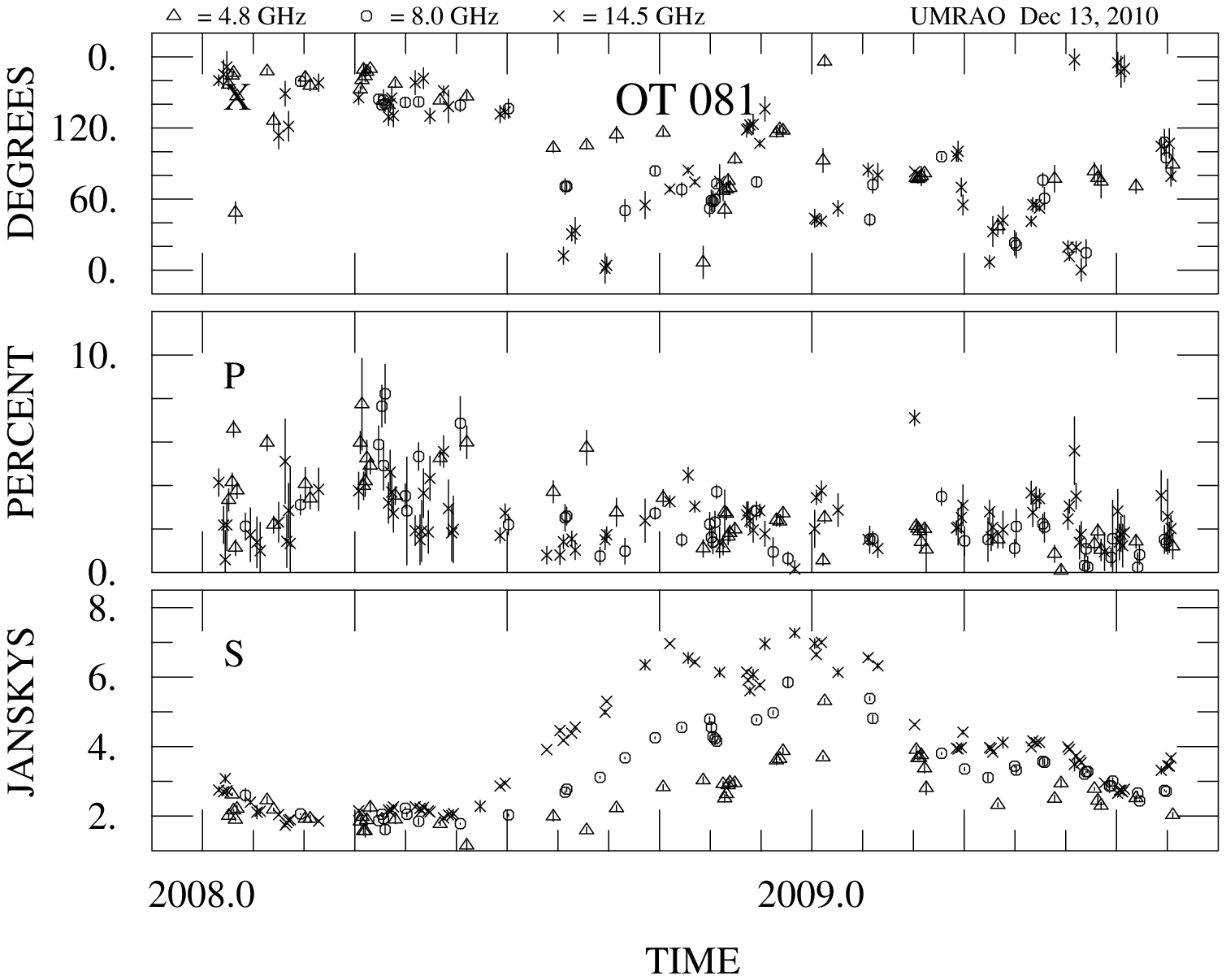}
\caption{From bottom to top: daily averages of the total flux density,
the fractional linear polarization, and the electric vector position angle
for OT~081 (1749+096). Symbols are as in Fig.~\ref{fig1}.}
\label{fig2}
\end{figure}

\begin{figure}
\plotone{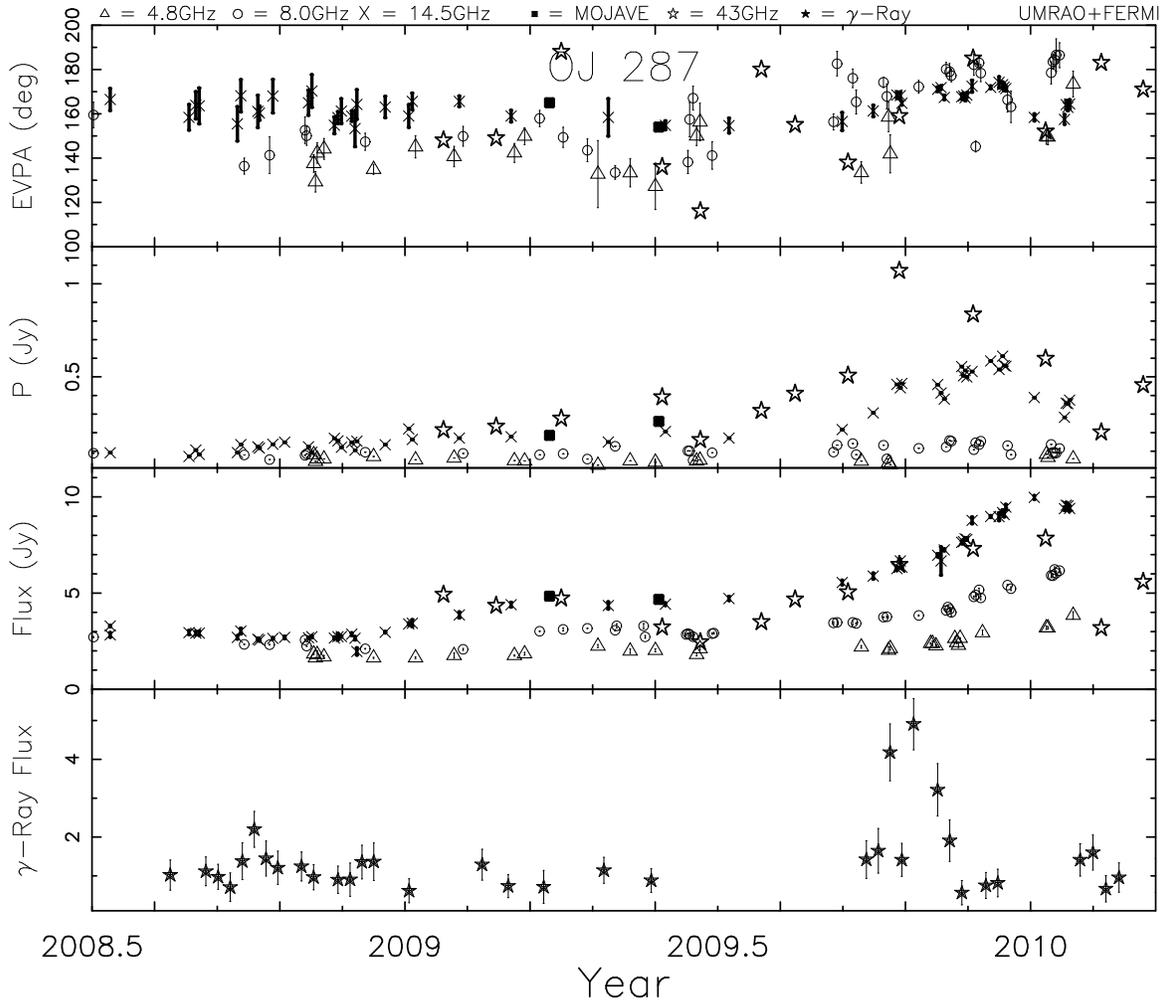}
\caption{From bottom to top: $\gamma$ ray light curve and daily averages
of the total flux density, the linearly polarized flux density, and
the electric vector position angle for OJ~287 (0851+202). Symbols are
as in Fig.~\ref{fig1} for the UMRAO monitoring data; source-integrated
VLBA data obtained from the MOJAVE (15~GHz) and BU (43~GHz) websites are
shown for comparison. The $\gamma$-ray light curve was kindly provided by
S. Jorstad. Units are photons/sec/cm$^2$x10$^{-7}$.}
\label{fig3}
\end{figure}

Generic properties of the variability in total flux density and linear
polarization are summarized in Table~\ref{table1}(A), for comparison with
models to be presented in \S\ref{results}. The values listed are based on
examination of the UMRAO database, while Figures~\ref{fig1}-\ref{fig3}
present specific examples illustrating the range in these properties
during recent well-resolved events temporally associated with $\gamma$-ray
flares detected by {\it Fermi}. The table contains: in column 2,  a simple
measure of the flare shape given by the rise time, i.e. the time from
start to peak, $\tau_{rise}$, divided by the event duration, T; in column
3 the spectral index at outburst start, at 14.5~GHz peak, and at end,
in the UMRAO frequency range 4.8 to 14.5~GHz; in column 4, characteristic
values of the maximum fractional linear polarization (note that while the
values listed are typical, values near 18\% have occasionally been observed
during flares); in column 5 the monotonic swing in EVPA during an event,
$\Delta$EVPA$_{time}$; and in column 6, the spectral variation in EVPA
throughout this swing, $\Delta$EVPA$_{freq}$.  The spectral indices follow
the sign convention $S_{\nu}\propto\nu^{+\alpha}$, and the values listed
for this parameter are based on numerical results for selected events,
using paired values of 2-week averages of the data at 14.5 and 4.8~GHz. For
linear polarization (peak fractional linear polarization and variation
in EVPA with time or frequency), the values listed were determined from
visual inspection of the long term UMRAO data. No corrections were applied
for Faraday rotation; but events in sources with low source-integrated
Faraday rotation measure were preferentially examined.

Figure~\ref{fig1} shows the UMRAO monitoring data for the QSO PKS~1510-089. A
series of $\gamma$-ray flares occurred during the time period shown which
have been analyzed in detail in \citet{abd10a} and in \citet{mar10}. A
particularly large and isolated event was identified with start date 10
March, 2009 and stop date 9 April, 2009 \citep{abd10a}. The UMRAO light
curves reveal that a radio band total flux density flare commenced in early
2009 at which time both a resolved linear polarization flare and an ordered
swing in EVPA are apparent at 14.5 and 8.0~GHz. Because of the differences
in variability time-scales in the radio and $\gamma$-ray bands and the
expected time delays produced by self-absorption in the emitting region,
there are some uncertainties in associating specific events across bands;
an event with the signature of an oblique shock, however, unambiguously
occurred during this time period which includes the $\gamma$-ray flare.

Figure~\ref{fig2} shows the radio band data during the period 2008.0 through
2009.5 for the very active BL~Lac object, OT~081 (1749+096), one of the
three sources originally modeled using the transverse shock formulation
\citep{hug91}. Several resolved events are apparent in both the total
flux density and fractional linear polarization light curves; an ordered
swing in EVPA at 14.5~GHz occurred from August to December 2008 followed
by a shorter-duration ordered swing during January-February 2009. The
Fermi light curve for this source is included in the variability study
based on data from the first year of {\it Fermi} operation \citep{abd10b}.
A very large flare in the $\gamma$-ray band was in progress in August 2008,
and a second flare which peaked in mid-March 2009 is also apparent in these
data. Activity, including strong flaring, occurred in both the $\gamma$-ray
and radio wavebands.

Figure~\ref{fig3} shows both the $\gamma$-ray and radio band light curves
for OJ~287. In this source the signature of a shock temporally associated
with the outburst observed by {\it Fermi} is clearly apparent. The swing
in EVPA through a limited range of about $40\arcdeg$ is consistent with
the passage of an oblique shock.

\section{Jet Model}\label{flow}
\subsection{The Quiescent Flow}\label{quiescent}
All computations are performed in a box $61\times61\times600$ cells in
extent, with the axis of the jet parallel to the long axis, $z$, and polar
angles are measured in the usual sense: $\theta$ from the $z$-direction,
and $\phi$ from the $x$-direction. A simple conical form has been taken for
the jet boundary as observations cannot constrain more complex profiles
explored over at most tens of jet radii. The opening angle ($\mu$) is
determined by setting the jet radius to be $r_{\rm lo}$ on the inflow
plane $z=0,$ and $r_{\rm hi}$ on the outflow plane $z=z_{\rm max}=600$;
angles are typically $1-2\arcdeg$.

Quiescent flow values are established across the entire domain, and then
an apodizing filter
\begin{equation}
a\left(\rho\right) = 0.5 \left[1+\tanh\left(\frac{-\rho+r\left(z\right)}{a_w}\right)\right],
\end{equation}
is applied, where $\rho=\left(x^2+y^2\right)^{1/2}$ is the radial
location of a cell at some $z$, where the jet radius is $r\left(z\right)$,
and $a_w$ is a parameter that controls the flow's boundary extent; $a_w$
is typically one cell.

Whether flows exhibit significant acceleration or deceleration
remains controversial, and indeed, might differ from source to source.
Deceleration has been inferred for FR 1 radio galaxies \citep {lai96,
lai99}, while a recent study of the BL Lac object OJ 287 finds that the
flow remains highly relativistic to distances as high as hundreds of
kiloparsecs from the nucleus \citep{mj10}.  In view of this uncertainty,
and the limited number of jet radii explored, the quiescent flow speed is
taken as a constant specified by its Lorentz factor, $\gamma_f$; values
are typically $2-5.$ A diverging flow is then established, with stream
lines parallel to the $z$-axis on axis, and parallel to the flow boundary
there. The magnetic field is constructed as described in \S\ref{magnetic},
and assumed to fall in strength along the flow as $r\left(z\right)^{-2}$.
\citet{leahy91} gives a good overview of the evolution of parameters in
a diverging adiabatic flow, and it might be thought that a fall-off as
$r\left(z\right)^{-4/3}$ would be appropriate for a flow of fixed speed,
given that the default assumption is that of a random field. However,
that assumes no coupling between the components, so that the component
perpendicular to the flow falls more slowly, driving the flow away from
isotropic turbulence.  The assumption of an isotropic random component
at every location in the quiescent flow implies a turbulent driving that
persists the length of the flow, transforming perpendicular into parallel
field, which declines with the assumed $r\left(z\right)^{-2}$ dependence,
and vice versa, coupling the decline of the perpendicular component to that
of the parallel component. Further, the helical ordered field that is also
explored, if generated from a turbulent component by a dissipative dynamo
process such as noted in \S\ref{ordered}, will then follow the same global
trend.  The adopted dependence is thus a simplification, but reasonable
given the other approximations.  As discussed in the next section, the
method used to generate a random field component, the ad hoc scaling along
the flow, and the scaling of the relative strengths of random and ordered
field components, mean that the field does not satisfy a self-consistent
magnetohydrodynamic model for its generation, and is not divergence-free.
However, the aim is to construct a plausible magnetic field topology with
which to explore the properties of the emergent radiation, not a field to
be evolved subject to constraints to ensure conservation and/or preservation
of the divergence-free character of the field.

The flow is assumed to be pervaded by a power law distribution of radiating
particles, with density
\begin{equation}
n\left(\gamma\right)\,d\gamma = n_0 \gamma^{-\delta}\,d\gamma,\ \ \ \gamma>\gamma_i,
\end{equation}
with the index $\delta$ fixed. Adiabatic gains and losses will not
change the slope, synchrotron losses are ignored over the tens of
jet radii propagation explored here, and no allowance is made for Fermi
acceleration at shocks, whose primary role will be to produce a small
number of particles radiating at frequencies above those probing the
quiescent emission. The absolute scaling of the density is arbitrary,
as functions of the density constant $n_0$, and the magnetic field
strength are absorbed into a fiducial optical depth, $\tau$, which acts
as a free parameter; see \S\ref{transfer}. The density constant, $n_0$,
is taken to fall due to adiabatic expansion of the flow according to
$r\left(z\right)^{-2\left(\delta+2\right)/3},$ and the cutoff thermal
Lorentz factor (see \S\ref{transfer} for the significance of this quantity),
$\gamma_i$, falls as $r\left(z\right)^{-2/3}.$

\subsection{Magnetic Field Structure}\label{magnetic}
Arguments in favor of a largely tangled magnetic field were made by
\citet{hug05}, the salient points being: a) the low degree of polarization
exhibited by compact extragalactic radio sources when in a quiescent
state has been widely interpreted as due to `root-N' depolarization
in a synchrotron source with many randomly oriented magnetic cells; b)
there have been successful models of the temporal, spatial, and spectral
attributes of outbursts in a number of individual sources, with a scenario
in which the shock compression of a flow provides an effective order to
the otherwise random magnetic field; c) the observed levels of linear
and circular polarization are best modeled by a scenario in which root-N
depolarization plays a significant role, albeit that some of the magnetic
energy is in an ordered component.

In contrast to this picture, Gabuzda and coworkers
\citep{gab04,gab05,mah08,osu08,con09} have argued in favor of a large-scale
magnetic field -- in particular with a helical character -- the key evidence
being rotation-measure gradients across jets, which are interpreted as due
to the systematic change in the line-of-sight component of the jet magnetic
field. Furthermore, it is almost invariably the case that `central engine'
models invoke an ordered magnetic field in the environment of a supermassive
black hole \citep[see, by way of example][]{kom07,mac07} and this order
(together with that due to shear) would be expected to be imprinted upon the
parsec-scale flow.  Recent work by \citet{bro10} and \citet{tay10} notes that
the presence of a helical field does not necessarily imply that a simple
monotonic variation in rotation measure across the flow will be observed,
and the resolution currently available precludes definitively establishing
the character of any ordered field within the jet. However, clearly there
is evidence consistent with such a field in a number of sources.

If an ordered field exists, and its sense is determined by the spin of the
central black hole, and/or the rotation of the associated accretion disk,
then the sign of the resultant circular polarization would be expected to
be a fixed attribute of any source. However, Stokes V monitoring does not
provide a simple picture; while, some sources exhibit such a constancy,
in others changes in the handedness of circular polarization with time
and frequency are quite unambiguous over durations of months to years,
e.g., \citet{all10}.  Also, evidence for a reversal in the direction
of the rotation-measure gradient across the jet of 1803+784 presented by
\citet{mah09}, at large distance from the core and away from the portion of
the jet most affected by finite beam size, are at odds with the results of
simulations by \citet{bro10}, who argue against the `magnetic tower' model
for a temporal change in the sign of the rotation measure.  The analysis of
all-Stokes polarization data for 3C~279 \citep{hom09} also does not find
support for the role of a helical field in the production of circularly
polarized emission.  Some resolution of these conflicting perspectives
may yet come from a more careful consideration of the emission location
(the jet spine) and the source of Faraday rotation (the jet sheath; the
dominant contribution to Faraday effects in the models of \citet{bro10}).
Until higher spatial resolution data become available, and simulations
can track the distribution of radiating particles, the topology of the
magnetic field remains uncertain, and a contentious issue, and so here
both random and ordered fields are considered.

\subsubsection{Random Field Component}\label{random}
An elegant method of establishing a random magnetic field is to select
random phases, and random amplitudes from a Rayleigh distribution, for
the Fourier transform of the vector potential, $\tilde{\bf A}\left({\bf
k}\right)$. The Fourier transform of the magnetic field is then $\tilde{\bf
B}\left({\bf k}\right) =i{\bf k}\times \tilde{\bf A}\left({\bf k}\right),$
and an inverse Fourier transform of this yields a magnetic field that
is guaranteed to satisfy the divergence-free constraint. (See, e.g.,
\citet{tri91} for a detailed exposition of this approach.) However, as
first noted in this context by \citet{tom88}, such a technique leads to
large-scale Fourier components that give rise to a small but significant
excess polarized flux density. An exploration of this in the current study
showed that unacceptable levels of polarized flux density were produced,
and that there is no simple way to mitigate that effect.

\citet{tom88} used a lattice of nested cells, each with randomly chosen
field orientation, and with the amplitude of the field components scaled
to produce an approximately Kolmogorov power spectrum. The goal of the
current study is such that there is a need to extensively explore a very
large parameter space, and thus use a 3D grid of limited resolution, in
order to keep individual runs to a modest time. Furthermore, a jet length
of many jet radii is required, because although the total flux density
falls rapidly after a propagating enhancement to the particle and field
densities passes the $\tau=1$ surface, a polarized flux density significantly
above the quiescent value persists for much longer. Thus there are too
few cells within the jet to effectively employ a multi-scale approach,
and simply assign randomly chosen field directions within each cell. In
principle the field should be advected with the flow, but such advection
cannot be handled self-consistently at oblique shocks, due to the simple
way that those structures are modeled (see \S\ref{oblique}), and as this
study is concerned with `macroscopic' behavior in the polarized emission,
not random fluctuations arising from flow structures lighting up different
realizations of random field, a static random field is built into the jet.

To reduce the computational overhead, multiple realizations of a random
field have not been generated for each parameter combination. However,
in \S\ref{orientation} radiation transfer is performed for a range of
azimuthal angles about a flow with propagating oblique shock. By symmetry the
results for azimuthal angles $225\arcdeg$, $270\arcdeg$, and $315\arcdeg$
are identical to results for $135\arcdeg$, $90\arcdeg$, and $45\arcdeg$
respectively, except that the different angles sample different projections
of the random field component, and so the differences in emergent radiation
for the three pairs of runs provide insight into variations to be expected
from different realizations.  At all azimuthal angles, total intensity
curves are identical, as are those for percentage polarization, but during
outburst there are small differences in EVPA as a function of frequency;
this is greatest at the first time shown, the quiescent state, because of
the low level of polarized emission then.

\subsubsection{Ordered Field Component}\label{ordered}
As discussed above, ordered, and in particular `helical', magnetic fields
have received much attention over the last decade. In order to include the
most physically plausible ordered field component, with the least number of
free parameters, a force-free, minimum energy configuration is adopted. On
the sub-parsec scale of interest here, the flow is far enough from the
central engine that local evolutionary effects should dominate over the
influence of the engine's ergosphere or inner accretion disk.

Indeed, long-term changes in the character of a large-scale magnetic field
are possible, if jets are subjected to turbulence and a dissipative dynamo
process. Turbulence could be driven throughout the body of the flow due to
their ultra-high Reynolds number.  Mean field dynamos in sheared, turbulent
flows are known to occur \citep{rk03}, and although the process has not been
explored in cylindrical geometry, it can be speculated that the mean field
then takes a form similar to that of a force-free flux tube.  This field
exists in what would be the asymptotic region discussed by \citet{mac06},
and its symmetry is thus unrelated to the angular momentum of the `central
engine'. Although the process of field reversal is different from that
exhibited by the solar-terrestrial field -- being stochastic rather than
periodic -- recent simulations \citep{you08} have shown such behavior,
and it would naturally explain the observed circular polarization `flips'.

Whether the field evolves due to dynamo action or relaxation, a likely
end point is a force-free configuration $\nabla\times {\bf B}=\sigma{\bf
B}$ \citep{eh91}. It has been shown by \citet{kc85} that in cylindrical
geometry, only the modes $m=0$ and $m=1$ contribute to the general solution
to the force-free equation in the minimum energy state. For a range of
jet and ambient medium parameters, and for any parameters leading to a
jet with small deviation from axisymmetry, only the $m=0$ mode need be
considered. Then, by the divergence-free constraint $\nabla\cdot{\bf B}=0$,
the axial wavenumber $k=0$.

In circular cylindrical coordinates,
$\left(\rho,\phi,z\right)$, this leads to a field configuration
\begin{equation}
{\bf B} = B_0\left(0,J_1\left(K\rho\right),J_0\left(K\rho\right)\right), 
\end{equation}
where the $J$ are Bessel functions of integer order, and $K$ is a constant
that must be determined.  The only physical boundary condition is that the
radial component of ${\bf B}$  vanish there,  but as this is identically
zero for the $m=0$ mode, this boundary condition provides no constraint. By
choosing $K=2.41/r\left(z\right)$, with $r\left(z\right)$ the jet radius,
the axial field goes through a null at the boundary \citep{lun50}.
While it is in principle permissible for the axial field to reverse sign
inside the jet, the adopted value prevents that, and is consistent with the
simple field topology selected by evolution to a force-free, minimum energy
state. This field is adopted `locally', in that the jet radius increases
due to flow divergence, but note that it is not built self-consistently
into the prescription of the magnetic field.

The constant $B_0$ is varied to model flow divergence and shock compression.
The strength of the random component is established by scaling so that $<{\bf
B}^2_{\rm ran}>$ is a specified multiple of the ordered field strength.
In the absence of a fully self-consistent magnetohydrodynamic model, it
is not clear whether the random field strength should be a multiple of
the {\it local} ordered field strength (corresponding, for example, to a
turbulent generation of the former from the latter by a fixed number of eddy
turn-overs); or spatially fixed for a given jet radius, a multiple of the
axial ordered field strength (corresponding, for example, to a turbulent
generation of the former to fixed multiple of the kinetic energy density
in flow turbulence). Both approaches have been tried, revealing that the
choice has no impact on the final results, merely changing the precise value
of $f^2={\bf B}^2_{\rm ord}/<{\bf B}^2_{\rm ran}>\sim 1$ where a transition
in the characteristic behavior of the polarized flux density is evident.

\subsection{Axial Magnetic Field}\label{axial}
\begin{figure}
\plotone{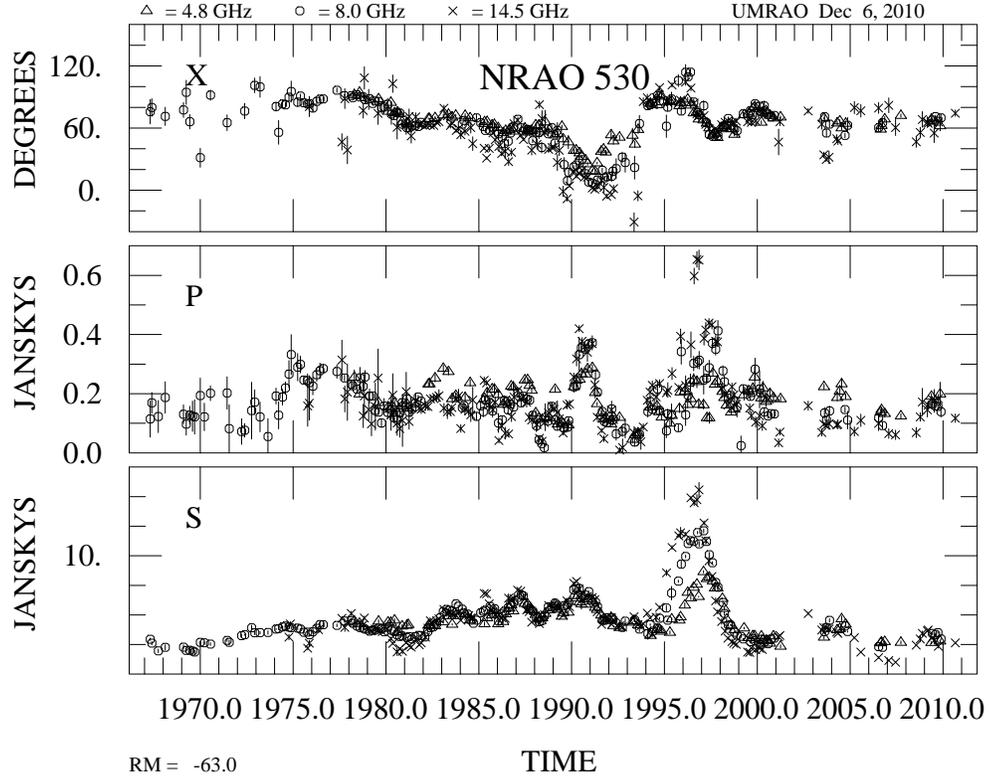}
\caption{From bottom to top:  60 day averages of the total flux density,
the linear polarization, and the electric vector position angle for
NRAO 530. Symbols are as in Fig.~\ref{fig1}. A rotation measure of -63
rad/m$^2$ has been assumed \citep{rus88}. The Faraday-corrected EVPA is
near $70\arcdeg$ at 14.5~GHz during the relatively quiescent period since
circa 2000 following a very large outburst in the 1990s. The position angle
of the jet determined from MOJAVE measurements at the same frequency is
$12\arcdeg$ \citep{kov05}.}
\label{fig4}
\end{figure}

It is extraordinarily difficult to define a quiescent state from data.
The UMRAO monitoring data exhibit almost continuous activity, and even
apparently inactive phases may be periods in which small-amplitude
outbursts are temporally unresolved. VLBI monitoring similarly reveals
few if any inactive epochs, and apparently-quiescent flows may contain
spatially unresolved components. The UMRAO database has been searched
to identify periods of `quiescence', during which there is measurable
linearly polarized emission. Between 1998 and 2001 0735+178 exhibited
a low level of constant flux density at the UMRAO frequencies, with
$\sim2$\% polarization and EVPA implying an ordered magnetic field lying
within some tens of degrees of the jet direction determined from MOJAVE
data \citep{lis09}. Similar numbers apply to NRAO~530 from 2002 to 2010 as
illustrated in Figure~\ref{fig4}.  Based on (admittedly sparse) examples such
as these, it is plausible to assume that in the quiescent state a weak axial
field is added to any random component, and in all simulations that follow,
an axial mean field with $2$\% the energy density of the random component
is added. This establishes a well-defined EVPA in the quiescent state.

\subsection{Oblique Shocks}\label{oblique}
Modeling a transverse structure comprising forward and reverse shocks,
separated by a contact discontinuity, is straightforward, as no lateral flow
is required: the shocked flow domain expands in the frame of the contact
surface, and either the expansion may be ignored during the brief interval
that the structure traverses the $\tau=1$ surface, or the expansion can be
modeled self-consistently from the jump conditions.  Furthermore, either
the forward or reverse shocked flow can be assumed to dominate emission,
or emission from both regions can be included.

A kinematic model for an oblique shock is difficult to construct, because
by its very nature, an oblique shock deflects the flow. Indeed, they have
long been invoked to model flow curvature in radio jets \citep{smi84}.
Deflection of the flow would appear to be inconsistent with the simple,
static, conical flow boundary assumed here. However, consider BL~Lac as
an example source: \citet{dmm00} present extensive monitoring by VLBI
which reveals knots with linear polarization behavior strongly suggestive
of oblique shocks. While these authors stress the curvature of the knot
trajectories, and difference between the kinematics of the four components
studied, an overlay of the trajectories (as defined by the data, rather
than the helical flow model) for components S7 and S9 exposes a remarkable
similarity, while the trajectories of S9 and S10 differ significantly
only at the last epoch of the later, and in that S10 may be followed
closer to the core. While component S8 does follow a distinctly different,
and more curved, trajectory, the general impression is of a well-defined
channel with modest curvature amplified in projection, that can support
the propagation of multiple oblique structures. Furthermore, studies such
as that by \citet{hug05} show that oblique shocks (which may be transient
features) can form and propagate, filling the cross section of a flow,
while that flow maintains a simple jet-like, rectilinear form.

An oblique shock is thus taken to be a plane spanning the jet, of obliquity
$\eta$, measured with respect to the direction of the upstream flow,
which is approximated as being in the sense of the jet axis, ignoring the
small divergence of the stream lines. $\eta=90\arcdeg$ corresponds to
a transverse shock, and the orientation of the shock is specified by the
azimuthal direction of the shock normal, $\psi$. The shocked flow is a volume
extending from this plane to a parallel plane, a distance $w$ along the flow
axis. The limit of the shocked flow is fixed; it is on the `core' side of
the shock plane for a ``forward'' shock (moving faster than the underlying
jet flow), and on the other side for a ``reverse'' shock (being overtaken by
the jet flow). The shocked flow may be terminated by adiabatic expansion of
the shocked material, radiative losses within the flow, or simply reflect
the distance the downstream flow has extended since formation. Adiabatic
expansion is not relevant in this context in the current model, as by
construction, the flow is constrained to lie within the quiescent jet
boundary. Radiation losses could have imposed a frequency-dependent length
on the downstream flow \citep{mar85} prior to evolution through the modeled
domain, but as explained in \S\ref{transfer}, the UMRAO observing frequencies
are close (being separated by $\sim\sqrt{3}$), and the length determined by
radiation losses goes as $\nu^{-1/2}$, so little frequency-dependence of the
structure is expected. For the sake of definitiveness, it is assumed that the
length of shocked flow is simply the evolved extent of the downstream flow,
which changes little during the brief propagation through the modeled domain.

The shocked flow is characterized by a compression $\kappa<1$, and following
\citet{cc90}, the upstream flow speed in the rest frame of the shock,
$\beta^{'}_u$ is computed:
\begin{eqnarray}
\beta^{'2}_u & = & \nonumber \\
 & & \{\kappa^2\sin^2\eta-\cos^2\eta-1+[(\kappa^2\sin^2\eta-\cos^2\eta-1)^2+ \nonumber \\
 & & 4\kappa^2(8\sin^2\eta+1)\sin^2\eta-4\cos^2\eta]^{\frac{1}{2}}\}/ \nonumber \\
 & & \{2[\kappa^2(8\sin^2\eta+1)\sin^2\eta-\cos^2\eta]\}.
\label{betaup}
\end{eqnarray}
From \citet{lb85} it follows that the downstream flow speed (also in the
shock frame) is given by
\begin{equation}
\beta^{'}_d = \frac
{\left[\left(1-\beta^{'2}_u\cos^2\eta\right)^2+9\beta^{'4}_u\cos^2\eta\sin^2\eta\right]^{\frac{1}{2}}}
{3\beta^{'}_u\sin\eta},
\label{betadp}
\end{equation}
and the flow is deflected by angle $\zeta$ given by
\begin{equation}
\tan\zeta = \frac
{\tan^2\eta\left(3\beta^{'2}_u-1\right)-\left(1-\beta^{'2}_u\right)}
{\tan\eta\left(\tan^2\eta+1+2\beta^{'2}_u\right)}.
\label{zeta}
\end{equation}
The flow direction within the shocked region is modified by $\zeta$, the
shock speed, $\beta_s$, is computed from the assumed quiescent flow speed
($\beta_f$) and the upstream speed in the shock frame ($\beta{'}_u$),
and the downstream flow speed $\beta^{'}_d$ is transformed into the
observer's frame, $\beta_d$.  Within the shocked region of the jet, the
particle density is increased by $\kappa^{-\left(\delta+2\right)/3}$, and
the magnetic field components projected on the shock plane are increased
by the  compression factor.

\subsection{Time Evolution}\label{evolution}
As noted in \S\ref{quiescent}, evidence for the acceleration or deceleration
of flows is ambiguous, and comes primarily from the observed deviations from
simple rectilinear motion displayed by a number of superluminal components.
It is difficult to distinguish between an acceleration or deceleration of
the bulk flow, and that of shock waves on the flow, in addition to which
acceleration or deceleration might be in direction only, due to a (modest)
curvature of the jet \citep[see for example][]{mar06}.

In the absence of compelling evidence for shock acceleration or deceleration,
the propagation of the shock is modeled by a simple constant displacement
of the shocked region discussed in \S\ref{oblique}; radiation transfer is
performed for the quiescent flow at the first time step, and then over
the remaining $N-1\sim10$ time steps the center of the shocked region
traverses the flow so that the shocked domain lies just within the inflow
jet volume at the second time step, and would have left the jet volume after
the last time step.  The corresponding physical time will be determined by
the shock speed ($\beta_s$), computed as described in \S\ref{oblique}, and
the physical length of the flow; as the latter is subject to an arbitrary
scaling, the time evolution is arbitrarily scaled to match typical outburst
periods seen in the UMRAO data.

\section{Radiation Transfer}\label{transfer}
The transfer of polarized radiation is performed using the method described
by \citet{hug05}, and is briefly recapped here. It is based on a formulation
for the transfer of polarized radiation through a diffuse plasma, allowing
for emission, absorption, the birefringence effects of Faraday rotation and
mode conversion (which can produce modest levels of circular polarization),
and relativistic aberration and boost, which has been described in detail
by a number of authors, and is compactly summarized by \citet{tom88}.

The observer lies at arbitrary polar angles ($\theta$,$\phi$), defined
in the conventional sense with respect to the Cartesian system used to
describe the kinematic model. An array of lines-of-sight is established,
using a preset density of lines along the longest axis of the projection of
the computational volume on the plane of the sky, with a commensurate number
orthogonal to that, to ensure equal resolution in the two directions. For
the results presented below, a resolution of $128$ pixels along the longest
axis was used. For each line-of-sight, the algorithm finds the most
distant cell, and radiative transfer is performed, cell to cell, until
the `near side' of the volume is exited. Within each cell, an aberrated
magnetic field direction is computed from the rest frame field, velocity,
and observer location, and is used with transfer coefficients modified by
the relevant Doppler boost. For a given epoch of `observation' cell values
corresponding to the appropriate retarded time should be used. However, as
this leads to an enormous increase in the computational overhead, and can
lead to simulated VLBI maps that are very difficult to interpret, because
of the problem of unambiguously associating map features with physical flow
structures, the radiation transfer is initially performed  without
retarded time effects included, and the latter will be considered only when
critical to understanding particular features of light curves or maps.
Simulations to justify this strategy are presented in \S\ref{delay}.

Radiation transfer is performed at dimensionless observing frequencies
of 0.6, 1.0, and 1.8125 corresponding to the UMRAO values of 4.8, 8.0 and
14.5~GHz. As described in \citet{hug89a} a fiducial thermal\footnote{This
does not imply a thermal distribution of particle energies.} Lorentz factor
($\gamma_c$; `thermal' implies the Lorentz factor associated with the random
motion of the emitting particles, as opposed to the bulk flow Lorentz factor)
is adopted, which is the energy of those particles radiating at the central
observing frequency in the average magnetic field for the jet volume at the
first time step of evolution. The value of this fiducial Lorentz factor
is not in itself significant, but model fitting to the data constrains
how far below this the power law electron energy distribution extends (to
$\gamma_i$), because it is the presence of these lower energy, but still
relativistic, particles that produce Faraday effects in the absence of a
`cold' particle distribution; adopting a plausible fiducial value thus
enables us to estimate the low energy cutoff to the power law distribution.

Prior to the radiation transfer, the optical depth for each cell must
be known; this is computed from the dimensionless line-of-sight length
through the cell, cell magnetic field, and cell particle density. This is
scaled with a single adjustable parameter chosen to produce some `target'
optical depth through the entire volume at the central observing frequency,
in the average magnetic field and particle density for the jet volume at
the first time step of evolution.  An initial `target' optical depth is
chosen in light of the spectral characteristics of the data being modeled;
the value is then adjusted to fine-tune the model fit. Given the actual
observing frequencies, the choice of fiducial Lorentz factor implies a
particular magnetic field strength; given that, and the optical depth
needed to reproduce the data, knowledge of the physical scale of the jet
implies a value for the particle density, or vice versa. However, as the
concern here is only with learning what such model fitting can say
about the topology and orientation of observer, shock, and magnetic field,
the implied field and particle densities are not explored.

\section{Model Results}\label{results}
\subsection{Role Of Time Delay}\label{delay}
As discussed in \S\ref{intro}, our primary goal is to `revalidate' the `shock
in jet' model, demonstrating that oblique shocks retain the temporal and
spectral characteristics of the total and polarized flux density behavior
of cm-band outbursts previously modeled with transverse structures, while
accommodating more complex behavior of the EVPA. As `proof of concept'
parameters are chosen similar to those used to model outbursts in
BL~Lac \citep{hug89b}, using a bulk flow of Lorentz factor $\gamma_f=2.5$
and compression $\kappa=0.7$, but with a shock oriented at $45\arcdeg$
to the flow axis, with the shock deflection in the plane of the sky;
see Run A in Table~\ref{table2}.

Assuming a forward shock ({\bf F}), implies the shock plane moving over the
underlying flow at Lorentz factor $6.7$, which when viewed at an angle to
the flow axis of $10\arcdeg$ would produce an apparent superluminal motion
of the leading edge of $\beta_{\rm app}\sim 6.5$. This is a significantly
faster shock speed than adopted in modeling the 1980s outbursts, but the
latter adopted a reverse ({\bf R}) shock model with the observer at a larger
angle to the flow sense, guided by the low apparent speeds reported in
the literature at the time. The current choice of parameters is guided
by the much higher apparent speeds reported in the more recent literature
\citep{lis09}, bearing in mind that over decades of activity the jet flow
direction can change, either forward or reverse shocks might be evident at
different epochs, and that shock events within a single source could exhibit
a range of obliquities. As shown in \S\ref{transverse}, it is possible
for oblique shocks to give rise to even higher percentage polarization,
and thus more prominent, events than a transverse shock with similar
parameters, for flows seen within tens of degrees of the line-of-sight.

Of the other parameters, optically thin spectral index is reasonably
well constrained by observation and has been fixed accordingly. Others,
such as flow opening angle and fiducial Lorentz factor, were picked on the
basis of typical values discussed in the literature over many decades,
to see if choice of such `plausible' values admits model light curves
with the same characteristics as those observed. Cutoff Lorentz factor,
shock width and optical depth (see \S\ref{transfer}), are adjusted in an
attempt to reproduce light curves looking most like the example UMRAO data.
The `order fraction' and orientations are subject to study in later sections.

\begin{figure}
\plotone{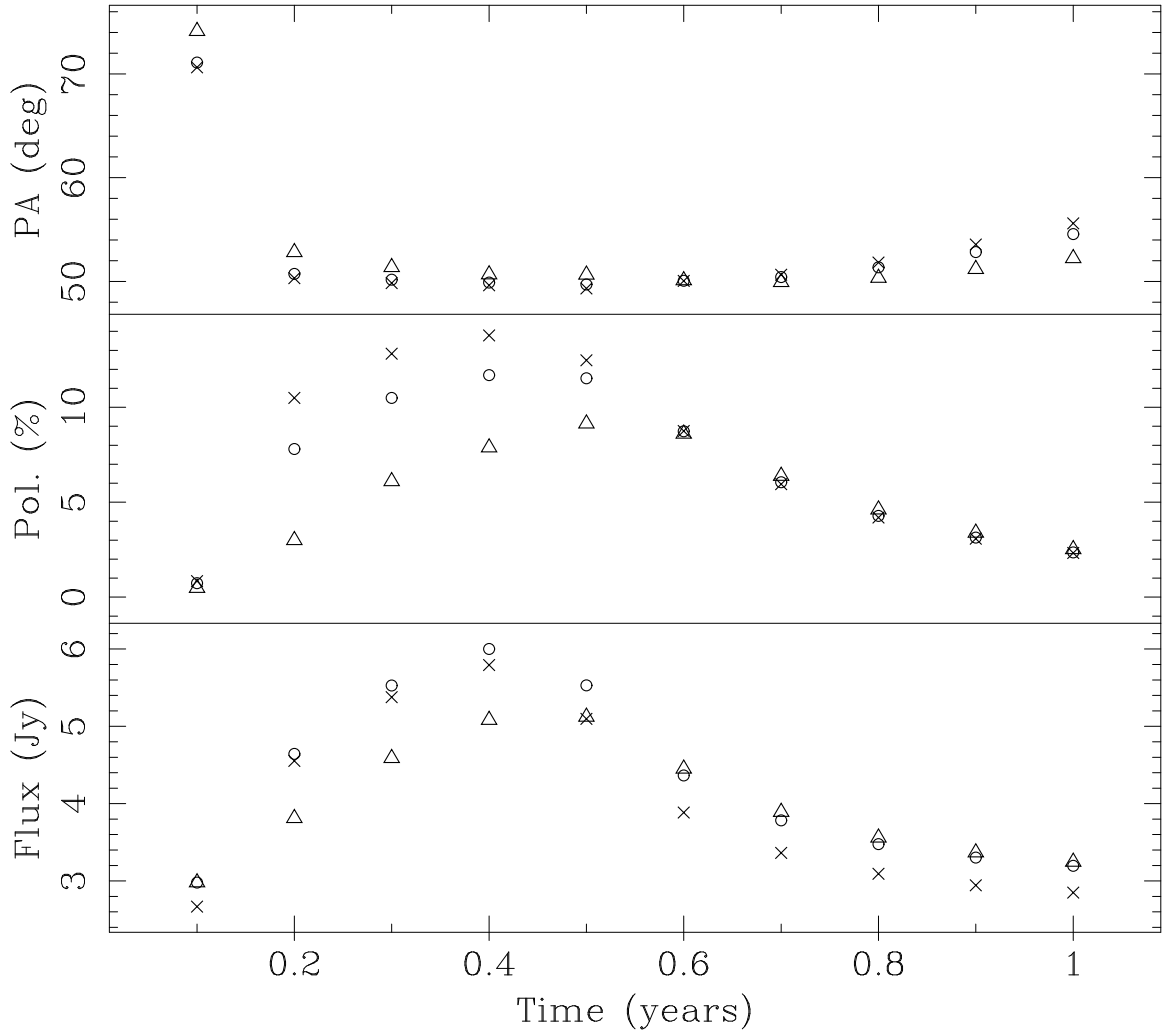}
\caption{Oblique shock evolution using the parameter set `Run A'
in Table~\ref{table2}. The panels and symbols correspond to those of
Fig.~\ref{fig1}, showing from bottom to top: the total flux density, the
fractional linear polarization, and the electric vector position angle,
at three frequencies corresponding to the UMRAO observations at 14.5,
8.0 and 4.8~GHz (crosses, circles and triangles respectively).}
\label{fig5}
\end{figure}

Figure~\ref{fig5} shows the evolution of flux density, percentage
polarization and EVPA for this model, without the inclusion of retarded
time effects. As described in \S\ref{transfer} the calculations are done
using dimensionless quantities; to `guide the eye', time and flux density
have been arbitrarily scaled to values representative of those seen in
the UMRAO data. The typical characteristics of UMRAO bursts described in
\S\ref{variability} -- the fractional flux density increase $\Delta S/<S>$,
spectral evolution through a partially optically thick phase, percentage
polarization with opacity/Faraday effects evident at the lowest frequency,
and swing in EVPA by tens of degrees -- are all reproduced. Note that
in earlier modeling it was found that to fit the spectral characteristics of
the polarized flux density, a fairly low cutoff thermal Lorentz factor
($\sim 20$) was needed for some sources. Here the general characteristics
of UMRAO outbursts are well-reproduced with a value that means opacity
effects dominate, with Faraday effects being only marginal: this implies
that in general few low energy electrons are present in these sources.

While use of retarded time in the modeling would be necessary for detailed
fits to data, the general characteristics of the total and polarized flux
density light curves and spectral behavior, even without using retarded
time, reproduce the behavior exhibited by the data (\S\ref{variability}).
Table~\ref{table1}(B) presents the model values corresponding to those
derived from the UMRAO database, and discussed in \S\ref{variability}.
A comparison of these is made in \S\ref{conclusions}.

\subsection{Orientation}\label{orientation}
Figure~\ref{fig6} shows models B1, B2, etc. -- the same model as in
Fig.~\ref{fig5} (see Table~\ref{table2}), but for a range of observer
orientation with respect to the shock plane. Recall that $\psi=0\arcdeg$, so
that the shock normal lies in the $x$-$z$ plane of the Cartesian coordinate
system. The observer orientations explored here ($0\arcdeg$, $45\arcdeg$,
$90\arcdeg$, $135\arcdeg$, $180\arcdeg$, $225\arcdeg$, $270\arcdeg$, and
$315\arcdeg$) correspond to starting with a view parallel to the $x$-axis,
and then rotating around the jet so that by panel (E) the observer is
`behind' the shock.  As one would expect, the total flux density light
curves are minimally changed by a change in azimuthal orientation. For
an observer orientation within tens of degrees of $\phi_{obs}=0\arcdeg$,
orientation and aberration conspire to provide a more nearly `face-on'
view of the shocked flow, and the percentage polarization is small. Indeed,
at $\phi_{obs}=0\arcdeg$ the polarized flux density from the shock region
cancels the small orthogonally polarized flux density associated with
the axial field, as evidenced by the varied behavior in EVPA in panel
(A). However, the percentage polarization is approaching 5\% by orientations
of $45$ and $335\arcdeg$, and exceeds 10\% at peak over a large range of
angles. The important conclusion is that azimuthal orientation does not
play a major role in the total and polarized flux density outburst light
curves; special conditions do not need to be invoked for polarizations of
this order to be seen, and (subject to flow speed and polar orientation)
most oblique structures will give rise to significant levels of polarized
flux density.  As the azimuthal angle does not play a significant role,
a value of $90\arcdeg$ is adopted in what follows.

\subsection{Oblique versus Transverse Shocks}\label{transverse}
In this section the run of percentage polarization with jet inclination
for the transverse case is established, as a measure by which to judge the
behavior in the oblique case; it is then shown that adopting an oblique
shock does not radically change the behavior of percentage polarization --
in fact leading to slightly higher values at some angles. This provides
further evidence that the `shock in jet' model survives the introduction
of oblique structures necessary to explain the temporal EVPA behavior seen
in the UMRAO data, and the evolution of features found in time sequences of
VLBI maps.

Figure~\ref{fig7} shows models C1, C2, etc.,  -- the same model as in
Fig.~\ref{fig5} (see Table~\ref{table2}), but contrasting transverse and
oblique shocks for a range of observer orientation with respect to the flow
axis. As orientation is changed, the free parameter $\tau$ is adjusted
to ensure a similar spectral behavior of the total flux density. In panels
(A-C) the shock is transverse, so the azimuthal location of the observer
plays no role in determining appearance, and the observer is viewing at
angles $20\arcdeg$, $40\arcdeg$, and $60\arcdeg$ respectively to the jet
axis. Note that with increasing angle the decline in percentage polarization
is rather slow. From \citet{hug85}, a flow with compression $\kappa=0.7$,
seen edge-on, and in the absence of opacity and Faraday effects, would be
expected to exhibit polarized emission $\sim25$\%, dropping to $\sim8$\%
at an angle of $50\arcdeg$. Given the modest Lorentz factor of the shocked
flow in the observer frame, $\sim3$, radiation from this angle {\it outside}
the critical cone of the flow (namely, at $140\arcdeg$ to the flow axis
in the flow frame) would be seen by the observer viewing at $50\arcdeg$
to the flow axis, an orientation spanned by panels (B) and (C) in the
figure. In the simple transverse case, for this level of compression,
quite high levels of polarization will be seen well beyond those values
of viewing angle usually adopted in blazar modeling.  Table~\ref{table1}(C)
presents the model values for Run C1, corresponding to those derived from
the UMRAO database, and discussed in \S\ref{variability}.  The only large
difference compared with the values presented in Table~\ref{table1}(B)
is in the jump in EVPA as expected, as well as in the spread in EVPA at
outburst end.

In panels (D-F) the original oblique shock is viewed at angles $30\arcdeg$,
$50\arcdeg$, and $70\arcdeg$ to the jet axis, for an azimuthal orientation of
$90\arcdeg$. (For an azimuthal orientation of $0\arcdeg$ the situation will
be approximately -- subject to flow deflection and different aberration --
as for the transverse case, modulo an angular offset.) It can be seen that
a high level of polarized emission persists to quite large angle from the
flow axis, a result related to the geometrical effect that as the shock
becomes more nearly parallel to the flow, rotation in polar angle has no
effect on the appearance of a flow seen initially nearly edge-on. As with
azimuthal orientation, it is concluded that special conditions do not need to
be invoked, and that the observed levels of percentage polarization can
be seen for oblique shocks for a significant range of observer orientation
with respect to the flow axis.

\subsection{Magnetic Field}\label{topology}
Figure~\ref{fig8} shows models D1, D2, etc. -- the same model as in
Fig.~\ref{fig5} (see Table~\ref{table2}), but for a range of `order
multiple', a measure of the importance of an ordered component of
magnetic field.  The values explored here ($0.1$, $0.3$, $1.0$, and $3.0$)
correspond to moving from an essentially random field to an almost totally
ordered field.  A fixed multiple at each point within the flow is assumed;
as discussed in \S\ref{ordered}, trials show that the details of how
the relative strength of these two components is modeled does not have a
significant influence on the results.

As noted by \citet{lyu05}, low levels of percentage polarization may result
even if the emitting volume contains wholly ordered magnetic field, and
that is evident in the values seen here outside of shock events -- significant
compared with the random field case, but low relative to the maximum
value for optically thin synchrotron radiation.  Careful choice of field
topology might reduce the value further, but as argued in \S\ref{ordered},
the adopted one appears to be the most physically plausible.

The nature of the outburst in percentage polarization differs markedly
between the case of polarization induced by compression of a random field,
as seen in panel (A) of Fig.~\ref{fig8}, and the case of polarization
induced by asymmetric compression of an ordered field, as seen in panel
(D). In the latter case the amplitude of the rise is limited (by less than
a factor $\sim 1.5$), with a strong frequency dependence unlike either
that seen in panel (A), or observed outbursts; the frequency dependence
arises because at the lowest frequency opacity causes polarized emission
to be dominated by counter-directed magnetic field arcs near to the jet
boundary. The behavior of the EVPA is also markedly different between
the cases. For the random field example the EVPA swings abruptly, and
similarly at all three frequencies, to an approximately constant value
during the long decline in outburst.  In the ordered field example there is
a strongly frequency-dependent decline, followed by an immediate recovery,
and no distinct plateau of EVPA during outburst. The evolution shown for
the cases of a predominantly ordered (helical) field neither matches what
has been seen in the evolution of the linearly polarized flux density during
well-resolved, distinct events that have been followed in detail, nor is
the spectral evolution of the position angle in the rise portion of the
event generally consistent with the observations, Based on this evidence,
it is concluded that the magnetic field within the emitting region cannot be
largely of the ordered helical type.

\section{Conclusions}\label{conclusions}
The `shock in jet' model for cm-waveband blazar variability has been
revisited, allowing for arbitrary shock orientation with respect to the
jet flow direction, and both random and ordered magnetic field. Oblique
shocks can explain events with swings in polarization position angle much
less than the $90\arcdeg$ associated with transverse structures, while
retaining the general characteristics of outbursts, including spectral
behavior and level of peak percentage polarization.

\begin{figure}
%emulate
\includegraphics[scale=0.9,clip=true]{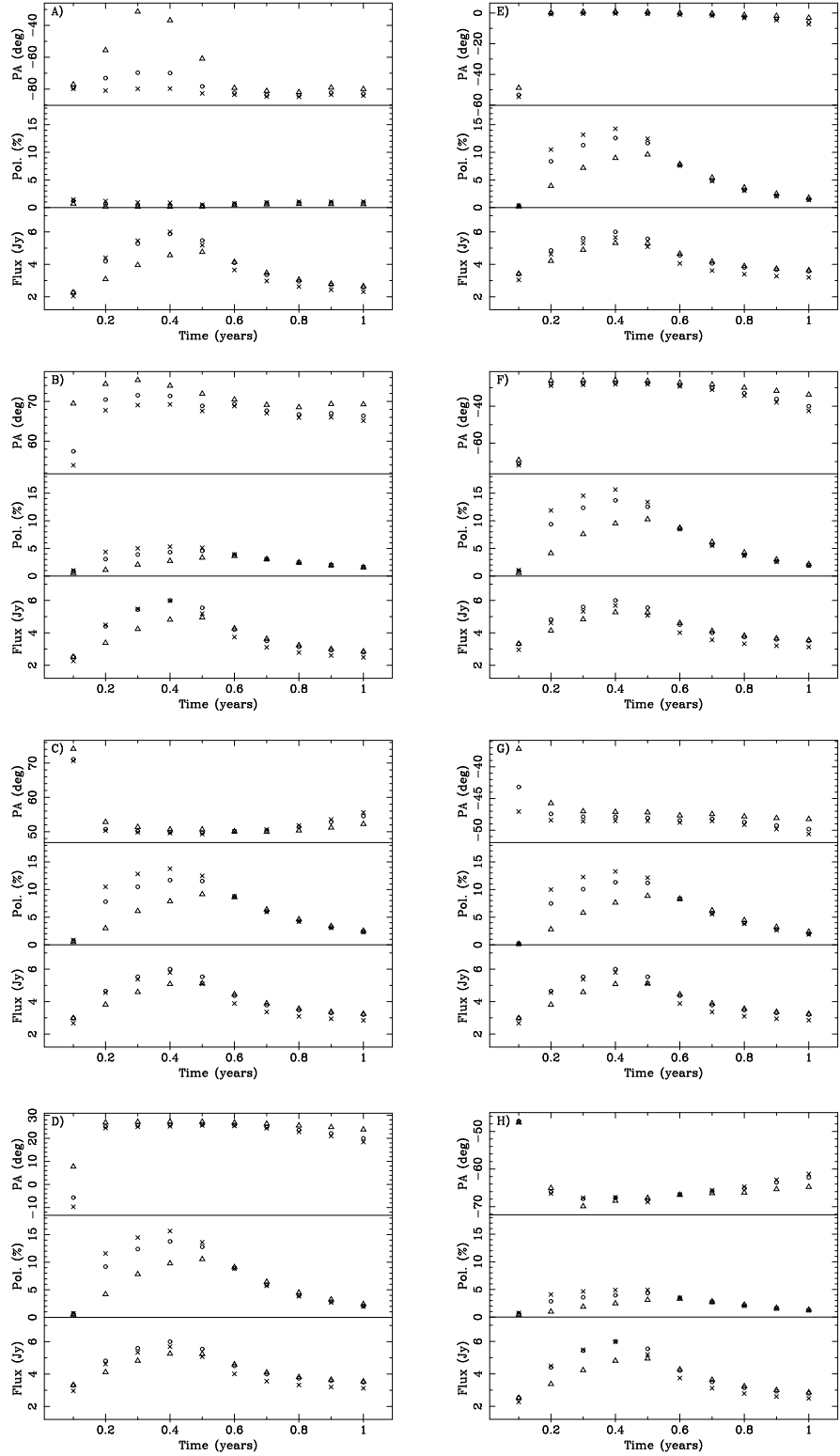}
%aastex
%\includegraphics[scale=0.8,clip=true]{fig6.ps}
\vskip -3.5cm
\caption{The same model as shown in Fig.~\ref{fig5}, for a range of observer
orientations with respect to the shock plane. From panel (A) the azimuthal
angle of the observer is $0\arcdeg$, $45\arcdeg$, $90\arcdeg$, $135\arcdeg$,
$180\arcdeg$, $225\arcdeg$, $270\arcdeg$, and $315\arcdeg$. (Note that in
these models the zero point for the EVPA is arbitrary and can change from
plot to plot -- only the {\it range} in values is significant.)}
\label{fig6}
\end{figure}

\begin{figure}
%emulate
\hskip -2.0cm \includegraphics[scale=1.0,clip=true]{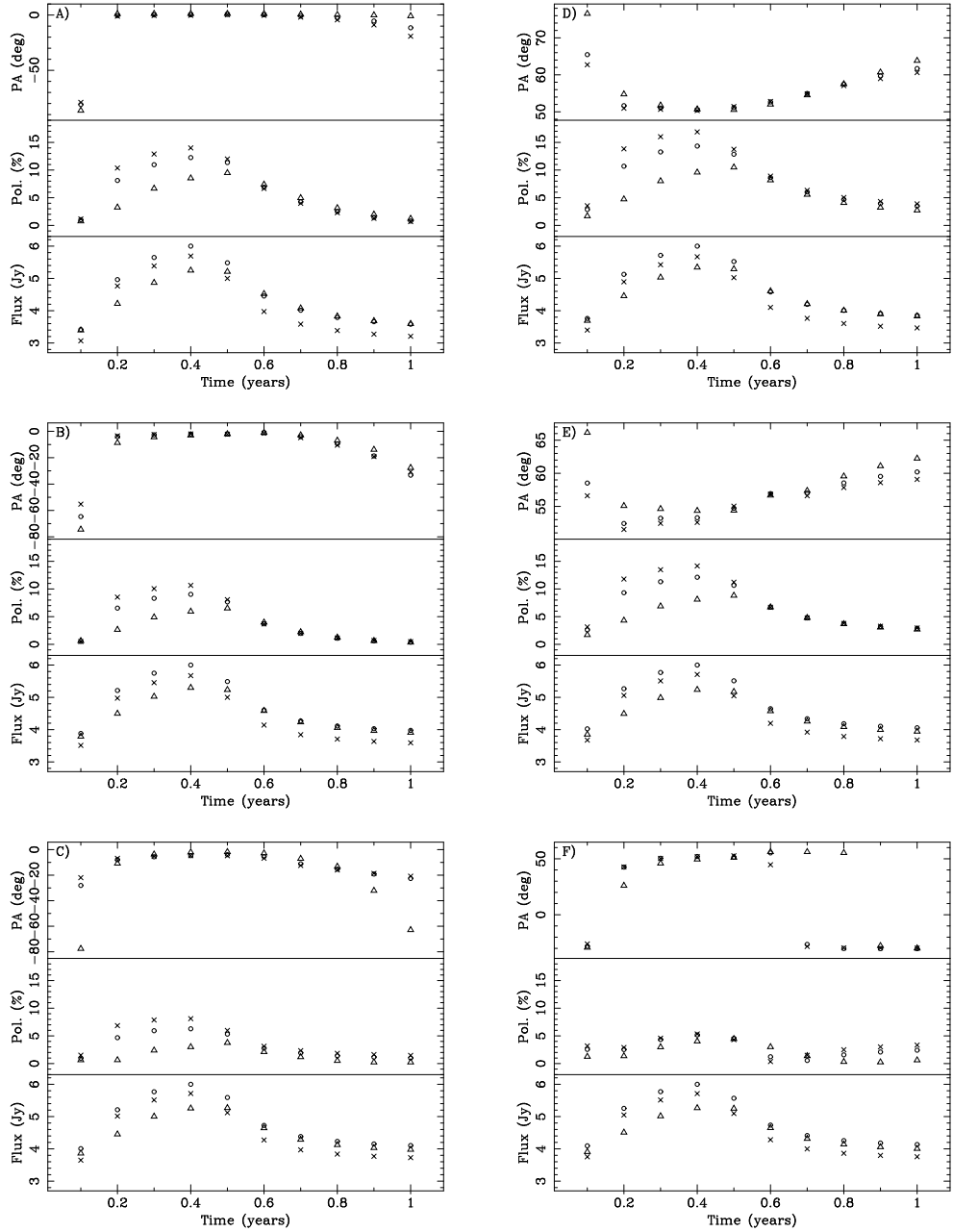}
%aastex
%\hskip -2.0cm \includegraphics[scale=0.9,clip=true]{fig7.ps}
\vskip -9.0cm
\caption{Comparison of a transverse shock model with that of oblique Run A,
as shown in Fig.~\ref{fig5}, for a range of polar angles. In the left panels
the shock orientation has been adjusted to be transverse to the flow, and in
panels (A-C) the observer is viewing at angles $20\arcdeg$, $40\arcdeg$,
and $60\arcdeg$ to the jet axis. The right panels are for the oblique
case, and in panels (D-F) the observer is viewing at angles $30\arcdeg$,
$50\arcdeg$, and $70\arcdeg$ to the jet axis.  (The range of angles is
chosen to optimally display variation with observer orientation. Note that
in these models the zero point for the EVPA is arbitrary and can change
from plot to plot -- only the {\it range} in values is significant.)}
\label{fig7}
\end{figure}

\begin{figure}
\hskip -5.0cm \includegraphics[scale=1.3,clip=true]{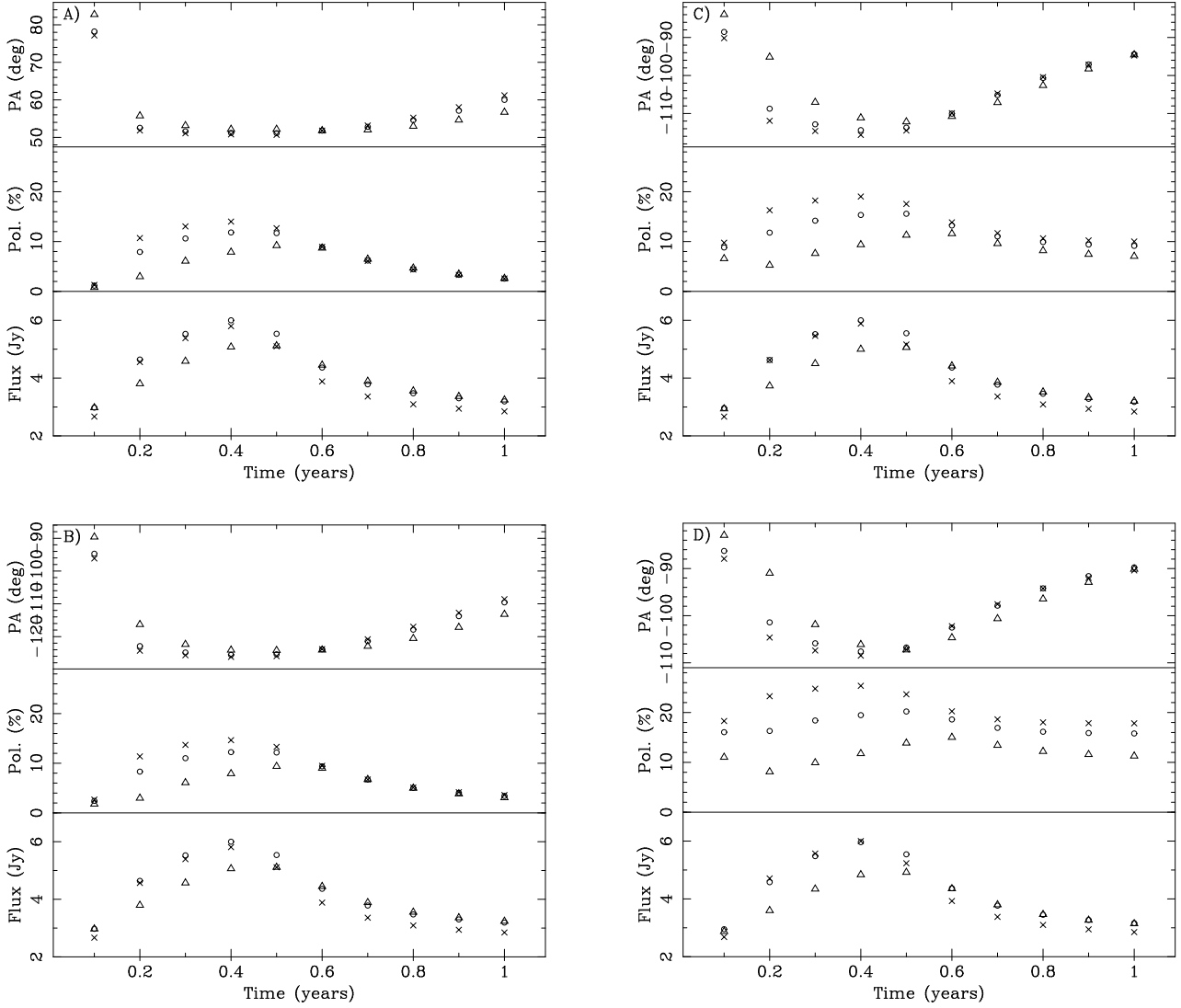}
\vskip -20.0cm
\caption{The same model as shown in Fig.~\ref{fig5}, for a range of
ordered to random magnetic field strengths. From panel (A) the `order
multiple' $f$ is $0.1$, $0.3$, $1.0$, and $3.0$, so that in panel (A)
the random field dominates, while in panel (D) the ordered field dominates.
(Note that in these models the zero point for the EVPA is arbitrary and can
change from plot to plot -- only the {\it range} in values is significant.)}
\label{fig8}
\end{figure}

Specifically, as can be seen from Table~\ref{table1}, the model reproduces
the temporal behavior of the UMRAO data: the trend in spectral evolution
through an opaque phase, while remaining quite flat throughout the burst
evolution; significant levels of percentage polarization (but far less
than the maximum permitted for a homogeneous source); the magnitude of
temporal evolution seen in EVPA; and the frequency spread in EVPA during
outburst of order degrees. The range of models provided by varying the
observer orientation with respect to the shock plane (\S\ref{orientation},
Fig.~\ref{fig6}) display the same general characteristics, but a range of
percentage polarization from almost zero to $\sim20$\%, corresponding
to the range seen in UMRAO data (\S\ref{variability}).

For certain azimuthal orientations with respect to the shock normal the
percentage polarization during outburst is low -- not much more than that
seen in the quiescent state -- but in general, levels of polarization
commensurate with that seen in UMRAO monitoring data occur, indicating
that oblique shocks can explain the data without recourse to special
observer orientation. Similarly, for the flow parameters adopted here
(modest Lorentz factor, leading to modest aberration) oblique shocks can
give rise to the observed levels of percentage polarization for a broad
range of orientation with respect to the flow axis, as is the case for a
transverse shock with similar parameters. As found by \citet{hug89b}, only
modeling of the detailed outburst profile for specific events, with the
inclusion of retarded time effects (in addition to matching the precise
level of polarization achieved at peak outburst) will better constrain
the viewing angle.

Models dominated by a force-free, minimum energy magnetic field configuration
(essentially helical) display a limited rise in percentage polarization, and
a frequency dependent swing in polarization position angle not in agreement
with the results of single-dish monitoring programs, implying that the
field cannot be dominated by an ordered component with the character of
a force-free flux rope in the quiescent state. This is in agreement with
the conclusions of a number of studies (discussed in \S\ref{magnetic})
that argue against a predominantly helical field configuration within the
emitting region.

Outbursts well-explained by the `shock in jet' model are present during
$\gamma$-ray flaring in several sources, supporting the idea that shock
events are associated with activity from the radio to $\gamma$-ray bands.
Some $\gamma$-ray flares might be associated not with propagating shocks,
but rather with propagating particle density/magnetic field enhancements that
encounter stationary structures such as a recollimation shock \citep{agu10}.
Such events are best explained if the magnetic field of the propagating
enhancement is random, and so our conclusion that the quiescent flow of
blazars is predominately random provides support for this explanation.

\acknowledgements
We thank the referee for a careful reading of the paper, and many detailed,
thoughtful, and constructive comments.  This work was made possible
by support from NSF grant NSF0607523, NASA Fermi grants NNX09AU16G and
NNX10AP16G, and by support for the operation of UMRAO from the University
of Michigan. This research has made use of data from the MOJAVE database
that is maintained by the MOJAVE team \citep{lis09}.

{}

\clearpage

\begin{deluxetable}{llllll}
\tablecolumns{6}
\tablewidth{0pt}
\tablecaption{Typical Flare Properties\label{table1}}
\tablehead{ & \colhead{$\tau_{rise}$/T} & \colhead{$\alpha$} & \colhead{P\%$_{max}$} & \colhead{$\Delta$EVPA$_{time}$} & \colhead{$\Delta$EVPA$_{freq}$}}
\startdata
\cutinhead{A. UMRAO Data}
 14.5~GHz  &     0.5  &     &  7   &                &         \\
  8.0      &    0.6        &     & 4    &           &             \\
  4.8      & 0.7   &    & 3.5  &                    &      \\
         &     &            &      &  $30-110\arcdeg$      &                   \\
  start  &     &  $+0.30$   &      &                       &  $20\arcdeg$      \\
  max   &      & $+0.5$  &      &          &  \\
  end    &      & $-0.2$  &      &          &    $0\arcdeg$   \\
\cutinhead{B. Shock Model A}
 14.5~GHz  &  0.4  &         & 14   &                 &      \\
  8.0      &  0.4  &         & 12   &                 &        \\
  4.8      &  0.5  &         &  9   &                 &      \\
         &       &           &      & $25\arcdeg$   &                 \\
  start  &       & $-0.11$   &      &               & $4\arcdeg$      \\
  max   &       &  $+0.13$  &      &              &  $1.5\arcdeg$  \\
  end    &      & $-0.13$   &      &                  &    $4\arcdeg$   \\
\cutinhead{C. Shock Model C1 ($\eta=90\arcdeg$, $\theta_{obs}=20\arcdeg$)}
14.5~GHz  &  0.4  &         & 14   &                 &              \\
 8.0             &  0.4  &         & 12.5   &               &
\\
 4.8             &  0.4  &       &  9.5   &               &              \\
          &       &         &        & $85\arcdeg$   &              \\
 start    &       & $-0.10$ &        &               & $7\arcdeg$   \\
 max      &       & $+0.07$ &        &               & $1.5\arcdeg$ \\
 end      &       & $-0.11$ &        &               & $19\arcdeg$   \\
\enddata 
\tablecomments{Specific values of $\alpha$ vary from event to event. Those
listed show a typical change during an observed flare.}
\end{deluxetable}

\begin{deluxetable}{llllll}
\tablewidth{0pt}
\tablecaption{Model Parameters\label{table2}}
\tablehead{
\colhead{Parameter} & \colhead{Run A} & \colhead{Runs B} & \colhead{Runs C} & \colhead{Runs D} }
% Run A: 002, 021, 044, 045 / 085
% Run B (orientation section): 046, 047, 045, 048, 049, 050, 051, 052 /
%                              086, 087, 085, 088, 089, 090, 091, 092
% Run C (polar section): 079, 081, 083, 073, 075, 077
% Run D (topology section): 054, 055, 059, 056 / 093, 094, 095, 096
\startdata
Semi-angle ($\mu$)        & $2.4\arcdeg$  &               &               & \\
Spectral index ($\alpha$) & $0.25$        &               &               & \\
Fiducial LF ($\gamma_c$)  & $1000.0$      &               &               & \\
Cutoff LF ($\gamma_i$)    & $50.0$        &               &               & \\
Order multiple ($f$)      &     --        &               &               & $0.1$, $0.3$, $1.0$, $3.0$\\
Bulk LF ($\gamma_f$)      & $2.5$         &               &               & \\
Sense                     & {\bf F}       &               &               & \\
Shock width ($w$)         & $0.4$         &               &               & \\
Compression ($\kappa$)    & $0.7$         &               &               & \\
Obliquity ($\eta$)        & $45\arcdeg$   &               & $45,90\arcdeg$   & \\
Orientation ($\psi$)      & $0\arcdeg$    &               &               & \\
Observer $\theta_{obs}$   & $10\arcdeg$   &               & $20,40,60\arcdeg$   & \\
                          &               &              &  $30,50,70\arcdeg$ & \\
Observer $\phi_{obs}$     & $90\arcdeg$   & $0\arcdeg$, $45\arcdeg$, $90\arcdeg$, $135\arcdeg$ &               & \\
                          &               & $180\arcdeg$, $225\arcdeg$, $270\arcdeg$, $315\arcdeg$ &               & \\
\enddata
\end{deluxetable}

\clearpage

\clearpage

\clearpage

\clearpage

\clearpage

\clearpage

\clearpage

\end{document}